\documentclass[12pt,a4paper]{article}
\usepackage{amsmath}
\usepackage{latexsym}
\makeatletter
\def\rddots{\mathinner{\mkern1mu\raise\p@%
    \vbox{\kern7\p@\hbox{.}}\mkern2mu%
    \raise4\p@\hbox{.}\mkern2mu\raise7\p@\hbox{.}\mkern1mu}}
\makeatother
\newcommand{\ket}[1]{{\vert{#1}\rangle}}

\newcommand{\fukuso}{{\mathbf C}}

\begin{document}

\title{\sl An Approximate Solution of the Jaynes--Cummings 
Model with Dissipation}
\author{
  Kazuyuki FUJII
  \thanks{E-mail address : fujii@yokohama-cu.ac.jp }\quad and\ \ 
  Tatsuo SUZUKI
  \thanks{E-mail address : suzukita@aoni.waseda.jp ; 
  i027110@sic.shibaura-it.ac.jp }\\
  ${}^{*}$Department of Mathematical Sciences\\
  Yokohama City University\\
  Yokohama, 236--0027\\
  Japan\\
  ${}^{\dagger}$Center for Educational Assistance\\
  Shibaura Institute of Technology\\
  Saitama, 337--8570\\
  Japan\\
  }
\date{}
\maketitle
\begin{abstract}
  In this paper we treat the Jaynes--Cummings model 
  with dissipation and give an approximate solution to the 
  master equation for the density operator {\bf under the general 
  setting} by making use of the Zassenhaus expansion.
\end{abstract}
%

%
%
%
%
In this paper we treat the Jaynes--Cummings model (\cite{JC}) with dissipation 
(or the quantum damped Jaynes--Cummings model in our terminology) and 
study the structure of general solution from a mathematical point of view 
in order to apply it to Quantum Computation or Quantum Control, which are 
our final target. As a general introduction to these topics see for example 
\cite{BP} and \cite{WS}. 

We believe that the model will become a good starting point to study more 
sophisticated models (with dissipation) in a near future, see for example 
{\bf Concluding Remarks} in the paper.

Let us start with the following phenomenological master equation 
for the density operator of the atom--cavity system 
in \cite{Scala et al-1} :
\begin{equation}
\label{eq:Q-D Jaynes-Cummings}
\frac{\partial}{\partial t}\rho=-i[H_{JC},\rho]
+
{\mu}
\left\{a\rho a^{\dagger}-\frac{1}{2}(a^{\dagger}a\rho+\rho a^{\dagger}a)\right\}
+
{\nu}
\left\{a^{\dagger}\rho{a}-\frac{1}{2}(aa^{\dagger}\rho+\rho aa^{\dagger})\right\}
\end{equation}
where $H_{JC}$ is the well--known Jaynes-Cummings Hamiltonian (see \cite{JC}) 
given by
\begin{eqnarray}
\label{eq:Jaynes-Cummings}
H_{JC}
&=&
\frac{\omega_{0}}{2}\sigma_{3}\otimes {\bf 1}+ 
\omega_{0}1_{2}\otimes a^{\dagger}a +
\Omega\left(\sigma_{+}\otimes a+\sigma_{-}\otimes a^{\dagger}\right)
\\
&=&
\left(
  \begin{array}{cc}
    \frac{\omega_{0}}{2}+\omega_{0}N & \Omega a             \\
    \Omega a^{\dagger} & -\frac{\omega_{0}}{2}+\omega_{0}N
  \end{array}
\right)  \nonumber
\end{eqnarray}
with
\[
\sigma_{+} = 
\left(
  \begin{array}{cc}
    0 & 1 \\
    0 & 0
  \end{array}
\right), \quad 
\sigma_{-} = 
\left(
  \begin{array}{cc}
    0 & 0 \\
    1 & 0
  \end{array}
\right), \quad 
\sigma_{3} = 
\left(
  \begin{array}{cc}
    1 & 0  \\
    0 & -1
  \end{array}
\right), \quad 
1_{2} = 
\left(
  \begin{array}{cc}
    1 & 0 \\
    0 & 1
  \end{array}
\right)
\]
, and $a$ and $a^{\dagger}$ are the annihilation and creation 
operators of an electro--magnetic field mode in a cavity, 
$N\equiv a^{\dagger}a$ is the number operator, and 
$\mu$ and $\nu$ ($\mu > \nu \geq 0$) are some constants 
depending on it (for example, a damping rate of the cavity mode). 

Note that the density operator $\rho$ is in $M(2;\fukuso)\otimes M({\cal F})=
M(2;M({\cal F}))$, namely
\begin{equation}
\label{eq:density operator}
\rho=
\left(
  \begin{array}{cc}
    \rho_{00} & \rho_{01}  \\
    \rho_{10} & \rho_{11}
  \end{array}
\right)
\in M(2;M({\cal F}))
\end{equation}
where $M({\cal F})$ is the set of all operators on the Fock space 
${\cal F}$ defined by
\begin{eqnarray*}
{\cal F}
&\equiv &\mbox{Vect}_{\fukuso}\{\ket{0},\ket{1},\ket{2},\ket{3},
\cdots \} \\
&=&
\left\{\sum_{n=0}^{\infty}c_{n}\ket{n}\ |\ \sum_{n=0}^{\infty}|c_{n}|^{2}<\infty\right\}
;\quad \ket{n}=\frac{(a^{\dagger})^{n}}{\sqrt{n!}}\ket{0}
\end{eqnarray*}
and ${\bf 1}$ is the identity operator.

We would like to solve (\ref{eq:Q-D Jaynes-Cummings}) explicitly. 
However, it is very hard to solve at the present time, so we must 
satisfy by giving some approximate solution to it. 

First of all we write down the equation (\ref{eq:Q-D Jaynes-Cummings}) 
in a component-wise manner. For that we set 
\begin{equation}
\label{eq:Jaynes-Cummings-2}
H_{JC}
=
\left(
  \begin{array}{cc}
    \frac{\omega_{0}}{2}+\omega_{0}N & \Omega a             \\
    \Omega a^{\dagger} & -\frac{\omega_{0}}{2}+\omega_{0}N
  \end{array}
\right)
\equiv
\left(
  \begin{array}{cc}
    A & B  \\
    C & D
  \end{array}
\right)
\end{equation}
for simplicity. 

\noindent
Then it is easy to see
\begin{eqnarray}
\label{eq:Each Component}
\dot{\rho}_{00}
&=&
-i(A\rho_{00}+B\rho_{10}-\rho_{00}A-\rho_{01}C)+  \nonumber \\
&&
{\mu}\left\{a\rho_{00}a^{\dagger}-
\frac{1}{2}(a^{\dagger}a\rho_{00}+\rho_{00}a^{\dagger}a)\right\}
+
{\nu}\left\{a^{\dagger}\rho_{00}{a}-
\frac{1}{2}(aa^{\dagger}\rho_{00}+\rho_{00}aa^{\dagger})\right\}, 
\nonumber \\
\dot{\rho}_{01}
&=&
-i(A\rho_{01}+B\rho_{11}-\rho_{00}B-\rho_{01}D)+  \nonumber \\
&&
{\mu}\left\{a\rho_{01}a^{\dagger}-
\frac{1}{2}(a^{\dagger}a\rho_{01}+\rho_{01}a^{\dagger}a)\right\}
+
{\nu}\left\{a^{\dagger}\rho_{01}{a}-
\frac{1}{2}(aa^{\dagger}\rho_{01}+\rho_{01}aa^{\dagger})\right\}, 
\nonumber \\
\dot{\rho}_{10}
&=&
-i(C\rho_{00}+D\rho_{10}-\rho_{10}A-\rho_{11}C)+ \nonumber \\
&&
{\mu}\left\{a\rho_{10}a^{\dagger}-
\frac{1}{2}(a^{\dagger}a\rho_{10}+\rho_{10}a^{\dagger}a)\right\}
+
{\nu}\left\{a^{\dagger}\rho_{10}{a}-
\frac{1}{2}(aa^{\dagger}\rho_{10}+\rho_{10}aa^{\dagger})\right\}, 
\nonumber \\
\dot{\rho}_{11}
&=&
-i(C\rho_{01}+D\rho_{11}-\rho_{10}B-\rho_{11}D)+ \nonumber \\
&&
{\mu}\left\{a\rho_{11}a^{\dagger}-
\frac{1}{2}(a^{\dagger}a\rho_{11}+\rho_{11}a^{\dagger}a)\right\}
+
{\nu}\left\{a^{\dagger}\rho_{11}{a}-
\frac{1}{2}(aa^{\dagger}\rho_{11}+\rho_{11}aa^{\dagger})\right\}
\end{eqnarray}
where $\dot{\rho}_{ij}=(\partial/\partial t)\rho_{ij}$ as usual.

Here let us remind some technique used in \cite{EFS} and \cite{FS}, 
which is very useful in some case. For a matrix $X=(x_{ij})\in M({\cal F})$ 
\[X=
\left(
\begin{array}{cccc}
x_{00} & x_{01} & x_{02} & \cdots  \\
x_{10} & x_{11} & x_{12} & \cdots  \\
x_{20} & x_{21} & x_{22} & \cdots  \\
\vdots & \vdots & \vdots & \ddots
\end{array}
\right)
\]
we correspond to the vector $\widehat{X}\in 
{\cal F}^{\mbox{dim}_{\fukuso}{\cal F}}$ as
\begin{equation}
\label{eq:correspondence}
X=(x_{ij})\ \longrightarrow\ 
\widehat{X}=(x_{00},x_{01},x_{02},\cdots;x_{10},x_{11},x_{12},\cdots;
x_{20},x_{21},x_{22},\cdots;\cdots \cdots)^{T}
\end{equation}
where $T$ means the transpose. Then the following formula
\begin{equation}
\label{eq:well--known formula}
\widehat{EXF}=(E\otimes F^{T})\widehat{X}
\end{equation}
holds for $E,F,X\in M({\cal F})$.

This and equations (\ref{eq:Each Component}) give
\begin{eqnarray}
\label{eq:as a Vector}
\dot{\hat{\rho}}_{00}
&=&
-i\left\{
\left(A\otimes {\bf 1}-{\bf 1}\otimes A^{T}\right)\hat{\rho}_{00}
-{\bf 1}\otimes C^{T}\hat{\rho}_{01}+B\otimes {\bf 1}\hat{\rho}_{10}
\right\}+  
\nonumber \\
&&
\left[
{\mu}
\left\{
a\otimes (a^{\dagger})^{T}-
\frac{1}{2}(a^{\dagger}a\otimes {\bf 1}+{\bf 1}\otimes a^{\dagger}a)
\right\}+
{\nu}
\left\{a^{\dagger}\otimes a^{T}-
\frac{1}{2}(aa^{\dagger}\otimes {\bf 1}+{\bf 1}\otimes aa^{\dagger})
\right\}
\right]
\hat{\rho}_{00},  \nonumber \\
&{}& \nonumber \\
\dot{\hat{\rho}}_{01}
&=&
-i\left\{
-{\bf 1}\otimes B^{T}\hat{\rho}_{00}
+\left(A\otimes {\bf 1}-{\bf 1}\otimes D^{T}\right)\hat{\rho}_{01}
+B\otimes {\bf 1}\hat{\rho}_{11}
\right\}+  
\nonumber \\
&&
\left[
{\mu}
\left\{
a\otimes (a^{\dagger})^{T}-
\frac{1}{2}(a^{\dagger}a\otimes {\bf 1}+{\bf 1}\otimes a^{\dagger}a)
\right\}+
{\nu}
\left\{a^{\dagger}\otimes a^{T}-
\frac{1}{2}(aa^{\dagger}\otimes {\bf 1}+{\bf 1}\otimes aa^{\dagger})
\right\}
\right]
\hat{\rho}_{01},  \nonumber \\
&{}& \nonumber \\
\dot{\hat{\rho}}_{10}
&=&
-i\left\{
C\otimes {\bf 1}\hat{\rho}_{00}+
\left(D\otimes {\bf 1}-{\bf 1}\otimes A^{T}\right)\hat{\rho}_{10}-
{\bf 1}\otimes C^{T}\hat{\rho}_{11}
\right\}+  
\nonumber \\
&&
\left[
{\mu}
\left\{
a\otimes (a^{\dagger})^{T}-
\frac{1}{2}(a^{\dagger}a\otimes {\bf 1}+{\bf 1}\otimes a^{\dagger}a)
\right\}+
{\nu}
\left\{a^{\dagger}\otimes a^{T}-
\frac{1}{2}(aa^{\dagger}\otimes {\bf 1}+{\bf 1}\otimes aa^{\dagger})
\right\}
\right]
\hat{\rho}_{10},  \nonumber \\
&{}& \nonumber \\
\dot{\hat{\rho}}_{11}
&=&
-i\left\{
C\otimes {\bf 1}\hat{\rho}_{01}-
{\bf 1}\otimes B^{T}\hat{\rho}_{10}
+\left(D\otimes {\bf 1}-{\bf 1}\otimes D^{T}\right)\hat{\rho}_{11}
\right\}+  
\nonumber \\
&&
\left[
{\mu}
\left\{
a\otimes (a^{\dagger})^{T}-
\frac{1}{2}(a^{\dagger}a\otimes {\bf 1}+{\bf 1}\otimes a^{\dagger}a)
\right\}+
{\nu}
\left\{a^{\dagger}\otimes a^{T}-
\frac{1}{2}(aa^{\dagger}\otimes {\bf 1}+{\bf 1}\otimes aa^{\dagger})
\right\}
\right]
\hat{\rho}_{11} \nonumber \\
&{}&
\end{eqnarray}
because ${\bf 1}$ and $N=a^{\dagger}a$ are diagonal (${\bf 1}^{T}=
{\bf 1},\ N^{T}=N$).

From (\ref{eq:density operator}) we set
\begin{equation}
\hat{\rho}=
\left(
\begin{array}{c}
\hat{\rho}_{00} \\
\hat{\rho}_{01} \\
\hat{\rho}_{10} \\
\hat{\rho}_{11} 
\end{array}
\right)\ \Longleftarrow \  
\rho=
\left(
  \begin{array}{cc}
    \rho_{00} & \rho_{01}  \\
    \rho_{10} & \rho_{11}
  \end{array}
\right),
\end{equation}
then we obtain the following ``canonical" form

\begin{eqnarray}
\label{eq:canonical form}
\frac{\partial}{\partial t}\hat{\rho}
=
&-i&
\left(
\begin{array}{cccc}
A\otimes {\bf 1}-{\bf 1}\otimes A^{T} & -{\bf 1}\otimes C^{T} & 
B\otimes {\bf 1} & 0 \\
-{\bf 1}\otimes B^{T} & A\otimes {\bf 1}-{\bf 1}\otimes D^{T} & 0 & 
B\otimes {\bf 1} \\
C\otimes {\bf 1} & 0 & 
D\otimes {\bf 1}-{\bf 1}\otimes A^{T} & -{\bf 1}\otimes C^{T} \\
0 & C\otimes {\bf 1} & -{\bf 1}\otimes B^{T} & 
D\otimes {\bf 1}-{\bf 1}\otimes D^{T}
\end{array}
\right)
\hat{\rho} \nonumber \\
&+&
\left(
\begin{array}{cccc}
L &    &    &     \\
   & L &    &     \\
   &    & L &     \\
   &    &    & L
\end{array}
\right)\hat{\rho}
\end{eqnarray}
with
\[
L=
{\mu}
\left\{
a\otimes (a^{\dagger})^{T}-
\frac{1}{2}(a^{\dagger}a\otimes {\bf 1}+{\bf 1}\otimes a^{\dagger}a)
\right\}+
{\nu}
\left\{a^{\dagger}\otimes a^{T}-
\frac{1}{2}(aa^{\dagger}\otimes {\bf 1}+{\bf 1}\otimes aa^{\dagger})
\right\},
\]
or more explicitly
\begin{footnotesize}
\begin{eqnarray}
&&\frac{\partial}{\partial t}\hat{\rho}=  \nonumber \\
&&
-i\left(
\begin{array}{cccc}
\omega_{0}(N\otimes {\bf 1}-{\bf 1}\otimes N) & -\Omega{\bf 1}\otimes (a^{\dagger})^{T} & 
\Omega a\otimes {\bf 1} & 0 \\
-\Omega{\bf 1}\otimes a^{T} & 
\omega_{0}+\omega_{0}(N\otimes {\bf 1}-{\bf 1}\otimes N) & 0 & \Omega a\otimes {\bf 1} \\
\Omega a^{\dagger}\otimes {\bf 1} & 0 & 
-\omega_{0}+\omega_{0}(N\otimes {\bf 1}-{\bf 1}\otimes N) & -\Omega{\bf 1}\otimes (a^{\dagger})^{T} \\
0 & \Omega a^{\dagger}\otimes {\bf 1} & -\Omega {\bf 1}\otimes a^{T} & 
\omega_{0}(N\otimes {\bf 1}-{\bf 1}\otimes N)
\end{array}
\right)
\hat{\rho} \nonumber \\
&&+
\left(
\begin{array}{cccc}
L &    &    &     \\
   & L &    &     \\
   &    & L &     \\
   &    &    & L
\end{array}
\right)\hat{\rho}
\end{eqnarray}
\end{footnotesize}
from (\ref{eq:Jaynes-Cummings-2}) :
\[
A=\frac{\omega_{0}}{2}+\omega_{0}N,\ B=\Omega a,\ 
C=\Omega a^{\dagger},\ D=-\frac{\omega_{0}}{2}+\omega_{0}N.
\]
Here a simplified notation $\omega_{0}$ in place of 
$\omega_{0}{\bf 1}\otimes {\bf 1}$ has been used.

\vspace{3mm}
Now we set
\begin{footnotesize}
\begin{eqnarray}
\label{eq:X}
&&X=
\left(
\begin{array}{cccc}
\omega_{0}(N\otimes {\bf 1}-{\bf 1}\otimes N) & -\Omega{\bf 1}\otimes (a^{\dagger})^{T} & 
\Omega a\otimes {\bf 1} & 0 \\
-\Omega{\bf 1}\otimes a^{T} & 
\omega_{0}+\omega_{0}(N\otimes {\bf 1}-{\bf 1}\otimes N) & 0 & \Omega a\otimes {\bf 1} \\
\Omega a^{\dagger}\otimes {\bf 1} & 0 & 
-\omega_{0}+\omega_{0}(N\otimes {\bf 1}-{\bf 1}\otimes N) & -\Omega{\bf 1}\otimes (a^{\dagger})^{T} \\
0 & \Omega a^{\dagger}\otimes {\bf 1} & -\Omega {\bf 1}\otimes a^{T} & 
\omega_{0}(N\otimes {\bf 1}-{\bf 1}\otimes N)
\end{array}
\right), \nonumber \\
&&{} \\
&&Y=
\label{eq:Y}
\left(
\begin{array}{cccc}
L &    &    &     \\
   & L &    &     \\
   &    & L &     \\
   &    &    & L
\end{array}
\right) 
\left(
L=
{\mu}
\left\{
a\otimes (a^{\dagger})^{T}-
\frac{1}{2}(a^{\dagger}a\otimes {\bf 1}+{\bf 1}\otimes a^{\dagger}a)
\right\}+
{\nu}
\left\{a^{\dagger}\otimes a^{T}-
\frac{1}{2}(aa^{\dagger}\otimes {\bf 1}+{\bf 1}\otimes aa^{\dagger})
\right\}
\right) \nonumber \\
&&{}
\end{eqnarray}
\end{footnotesize}
for simplicity.

Next, in order to rewrite $X$ and $Y$ in terms of Lie algebraic 
notations used in \cite{EFS} we set
\begin{eqnarray}
K_{+}&=&a^{\dagger}\otimes a^{T},\ \ 
K_{-}=a\otimes (a^{\dagger})^{T},\ \ 
K_{3}=\frac{1}{2}(N\otimes {\bf 1}+{\bf 1}\otimes N+{\bf 1}\otimes {\bf 1}), 
\nonumber \\
K_{0}&=&N\otimes {\bf 1}-{\bf 1}\otimes N.
\end{eqnarray}
Then it is easy to see
\begin{eqnarray}
\label{eq:relations I}
&&(K_{+})^{\dagger}=K_{-},\ \ (K_{3})^{\dagger}=K_{3}, 
\ \ (K_{0})^{\dagger}=K_{0}, \nonumber \\
&&[K_{3},K_{+}]=K_{+},\ \ 
[K_{3},K_{-}]=-K_{-},\ \ 
[K_{+},K_{-}]=-2K_{3}, \nonumber \\
&&[K_{0},K_{+}]=[K_{0},K_{-}]=[K_{0},K_{3}]=0
\end{eqnarray}
and
\begin{eqnarray}
\label{eq:relations II}
&&[K_{0},a\otimes {\bf 1}]+a\otimes {\bf 1}=0,\ \ 
[K_{0},{\bf 1}\otimes a^{T}]+{\bf 1}\otimes a^{T}=0, \nonumber \\
&&[K_{0},a^{\dagger}\otimes {\bf 1}]-a^{\dagger}\otimes {\bf 1}=0,\ \ 
[K_{0},{\bf 1}\otimes (a^{\dagger})^{T}]-{\bf 1}\otimes (a^{\dagger})^{T}=0.
\end{eqnarray}
Namely, $\{K_{+},K_{-},K_{3}\}$ are generators of the Lie algebra 
$su(1,1)$, see for example \cite{KF} as a general introduction.

Under the above notations $X$ and $Y$ in (\ref{eq:X}) and (\ref{eq:Y}) 
can be rewritten as

\begin{equation}
\label{eq:X modified}
X=
\left(
\begin{array}{cccc}
\omega_{0}K_{0} & -\Omega{\bf 1}\otimes (a^{\dagger})^{T} & \Omega a\otimes {\bf 1} & 0 \\
-\Omega{\bf 1}\otimes a^{T} & \omega_{0}+\omega_{0}K_{0} & 0 & \Omega a\otimes {\bf 1} \\
\Omega a^{\dagger}\otimes {\bf 1} & 0 & -\omega_{0}+\omega_{0}K_{0} & -\Omega{\bf 1}\otimes (a^{\dagger})^{T} \\
0 & \Omega a^{\dagger}\otimes {\bf 1} & -\Omega {\bf 1}\otimes a^{T} & \omega_{0}K_{0}
\end{array}
\right)
\end{equation}
and
\begin{eqnarray}
\label{eq:Y modified}
Y
&=&
\left(
\begin{array}{cccc}
L &    &    &     \\
   & L &    &     \\
   &    & L &     \\
   &    &    & L
\end{array}
\right)\ \ 
\left(
L=
\nu K_{+}+\mu K_{-}-(\mu+\nu)K_{3}+\frac{\mu-\nu}{2}
\right) \nonumber \\
&\equiv&
\frac{\mu-\nu}{2}+\left(\nu K_{+}+\mu K_{-}-(\mu+\nu)K_{3}\right)1_{4}
\quad (\mbox{for simplicity})
\end{eqnarray}
where in the process the relation $aa^{\dagger}=a^{\dagger}a+{\bf 1}=N+{\bf 1}$ 
has been used.

As a result we obtain the final form
\begin{equation}
\label{eq:final form}
\frac{\partial}{\partial t}\hat{\rho}=(-iX+Y)\hat{\rho}
\end{equation}
with $X$ in (\ref{eq:X modified}) and $Y$ in (\ref{eq:Y modified}). This is 
our main result and we believe that the equation is clear--cut enough (compare 
it with (\ref{eq:Q-D Jaynes-Cummings})).  

Since the general solution is given by
\begin{equation}
\label{eq:general solution}
\hat{\rho}(t)=e^{t(-iX+Y)}\hat{\rho}(0)
\end{equation}
in a formal way we must calculate the term\ $e^{t(-iX+Y)}$, 
which is in general very hard. 
For that the following Zassenhaus formula is convenient.

\vspace{3mm}\noindent
{\bf Zassenhaus Formula}\ \ We have an expansion
\begin{equation}
\label{eq:Zassenhaus formula}
e^{t(A+B)}=
\cdots 
e^{-\frac{t^{3}}{6}\{2[[A,B],B]+[[A,B],A]\}}
e^{\frac{t^{2}}{2}[A,B]}
e^{tB}
e^{tA}.
\end{equation}
The formula is a bit different from that of \cite{CZ}. 

\vspace{3mm}
In this paper we use the approximation
\[
e^{t(A+B)}\approx e^{\frac{t^{2}}{2}[A,B]}e^{tB}e^{tA}
\]
with $A=-iX$ and $B=Y$, so that  $\hat{\rho}(t)$ is 
approximated as
\begin{equation}
\label{eq:approximate formula}
\hat{\rho}(t)\approx e^{-i\frac{t^{2}}{2}[X,Y]}e^{tY}e^{-itX}
\hat{\rho}(0).
\end{equation}

Let us calculate each term explicitly :

\vspace{5mm}\noindent
[I]\ \ First, we calculate $e^{-itX}$. However, the calculation 
is more or less well--known. We decompose $X$ into two parts
\begin{eqnarray}
X_{1}&=&
\left(
\begin{array}{cccc}
\omega_{0}K_{0} & -\Omega{\bf 1}\otimes (a^{\dagger})^{T} & 0 & 0 \\
-\Omega{\bf 1}\otimes a^{T} & \omega_{0}+\omega_{0}K_{0} & 0 & 0 \\
0 & 0 & -\omega_{0}+\omega_{0}K_{0} & -\Omega{\bf 1}\otimes (a^{\dagger})^{T} \\
0 & 0 & -\Omega {\bf 1}\otimes a^{T} & \omega_{0}K_{0}
\end{array}
\right), \nonumber \\
X_{2}&=&
\left(
\begin{array}{cccc}
0 & 0 & \Omega a\otimes {\bf 1} & 0 \\
0 & 0 & 0 & \Omega a\otimes {\bf 1} \\
\Omega a^{\dagger}\otimes {\bf 1} & 0 & 0 & 0 \\
0 & \Omega a^{\dagger}\otimes {\bf 1} & 0 & 0
\end{array}
\right).
\end{eqnarray}
Then it is not difficult to check
\begin{equation}
[X_{1},X_{2}]=0
\end{equation}
owing to the relations (\ref{eq:relations II}). If we set
\begin{equation}
\label{eq:swap}
S=
\left(
\begin{array}{cccc}
1 & 0 & 0 & 0  \\
0 & 0 & 1 & 0  \\
0 & 1 & 0 & 0  \\
0 & 0 & 0 & 1
\end{array}
\right)
\end{equation}
then
\begin{equation}
SX_{2}S^{-1}=SX_{2}S
=\Omega
\left(
\begin{array}{cccc}
0 & a\otimes {\bf 1} & 0 & 0 \\
a^{\dagger}\otimes {\bf 1}& 0 & 0 & 0 \\
0 & 0 & 0 & a\otimes {\bf 1} \\
0 & 0 & a^{\dagger}\otimes {\bf 1} & 0
\end{array}
\right).
\end{equation}
Moreover, if we decompose like
\[
\left(
\begin{array}{cc}
\omega_{0}K_{0} & -\Omega{\bf 1}\otimes (a^{\dagger})^{T}  \\
-\Omega{\bf 1}\otimes a^{T} & \omega_{0}+\omega_{0}K_{0}
\end{array}
\right)
=
\left(
\begin{array}{cc}
\omega_{0}K_{0} & 0                 \\
0 & \omega_{0}+\omega_{0}K_{0}
\end{array}
\right)
+
\left(
\begin{array}{cc}
0 & -\Omega{\bf 1}\otimes (a^{\dagger})^{T}  \\
-\Omega{\bf 1}\otimes a^{T} & 0
\end{array}
\right)
\]
then
\[
\left[
\left(
\begin{array}{cc}
\omega_{0}K_{0} & 0                 \\
0 & \omega_{0}+\omega_{0}K_{0}
\end{array}
\right),
\left(
\begin{array}{cc}
0 & -\Omega{\bf 1}\otimes (a^{\dagger})^{T}  \\
-\Omega{\bf 1}\otimes a^{T} & 0
\end{array}
\right)
\right]
=
0
\]
owing to the relations (\ref{eq:relations II}). Therefore 
we have a decomposition
\begin{eqnarray}
X_{1}&=&
\left(
\begin{array}{cccc}
\omega_{0}K_{0} & 0 & 0 & 0 \\
0 & \omega_{0}+\omega_{0}K_{0} & 0 & 0 \\
0 & 0 & -\omega_{0}+\omega_{0}K_{0} & 0 \\
0 & 0 & 0 & \omega_{0}K_{0}
\end{array}
\right)- \nonumber \\
&\Omega&
\left(
\begin{array}{cccc}
0 & {\bf 1}\otimes (a^{\dagger})^{T} & 0 & 0 \\
{\bf 1}\otimes a^{T} & 0 & 0 & 0 \\
0 & 0 & 0 & {\bf 1}\otimes (a^{\dagger})^{T} \\
0 & 0 & {\bf 1}\otimes a^{T} & 0
\end{array}
\right).
\end{eqnarray}
As a result we have a decomposition consisting of three 
commutative operators
\begin{eqnarray}
\label{eq:decomposition I}
X&=&
\left(
\begin{array}{cccc}
\omega_{0}K_{0} & 0 & 0 & 0 \\
0 & \omega_{0}+\omega_{0}K_{0} & 0 & 0 \\
0 & 0 & -\omega_{0}+\omega_{0}K_{0} & 0 \\
0 & 0 & 0 & \omega_{0}K_{0}
\end{array}
\right)-  \nonumber \\
&\Omega&
\left(
\begin{array}{cccc}
0 & {\bf 1}\otimes (a^{\dagger})^{T} & 0 & 0 \\
{\bf 1}\otimes a^{T} & 0 & 0 & 0 \\
0 & 0 & 0 & {\bf 1}\otimes (a^{\dagger})^{T} \\
0 & 0 & {\bf 1}\otimes a^{T} & 0
\end{array}
\right)+  \nonumber \\
&\Omega&
S
\left(
\begin{array}{cccc}
0 & a\otimes {\bf 1} & 0 & 0 \\
a^{\dagger}\otimes {\bf 1}& 0 & 0 & 0 \\
0 & 0 & 0 & a\otimes {\bf 1} \\
0 & 0 & a^{\dagger}\otimes {\bf 1} & 0
\end{array}
\right)
S.
\end{eqnarray}

\vspace{10mm}
Now, by making use of the well--known formula
\begin{equation}
\mbox{exp}
\left(
-i\alpha
\left(
\begin{array}{cc}
  0             & A  \\
A^{\dagger} & 0
\end{array}
\right)
\right)
=
\left(
\begin{array}{cc}
\cos(\alpha\sqrt{AA^{\dagger}}) & 
-i\frac{1}{\sqrt{AA^{\dagger}}}\sin(\alpha\sqrt{AA^{\dagger}})A   \\
-i\frac{1}{\sqrt{A^{\dagger}A}}\sin(\alpha\sqrt{A^{\dagger}A})A^{\dagger} & 
\cos(\alpha\sqrt{A^{\dagger}A}) 
\end{array}
\right)
\end{equation}
where $\alpha$ is some parameter, we can calculate $e^{-itX}$ easily. 
The result is
\begin{equation}
\label{eq:I}
e^{-itX}
=
\left(
\begin{array}{cccc}
  (11) & (12) & (13) & (14)  \\
  (21) & (22) & (23) & (24)  \\
  (31) & (32) & (33) & (34)  \\
  (41) & (42) & (43) & (44)
\end{array}
\right)
\end{equation}
where for $\alpha=\Omega t$ 
\begin{eqnarray*}
(11)
&=&e^{-it\omega_{0}K_{0}}
\cos(\alpha\sqrt{{\bf 1}\otimes aa^{\dagger}})
\cos(\alpha\sqrt{aa^{\dagger}\otimes {\bf 1}}) \\
&=&e^{-it\omega_{0}N}\cos(\alpha\sqrt{N+1})\otimes
e^{it\omega_{0}N}\cos(\alpha\sqrt{N+1}), \\
(12)
&=&i\ e^{-it\omega_{0}K_{0}}
\frac{1}{\sqrt{{\bf 1}\otimes aa^{\dagger}}}
\sin(\alpha\sqrt{{\bf 1}\otimes aa^{\dagger}})({\bf 1}\otimes (a^{\dagger})^{T})
\cos(\alpha\sqrt{aa^{\dagger}\otimes {\bf 1}}) \\
&=&i\ e^{-it\omega_{0}N}\cos(\alpha\sqrt{N+1})\otimes
e^{it\omega_{0}N}\frac{1}{\sqrt{N+1}}\sin(\alpha\sqrt{N+1})(a^{\dagger})^{T}, \\
(13)
&=&-i\ e^{-it\omega_{0}K_{0}}
\cos(\alpha\sqrt{{\bf 1}\otimes aa^{\dagger}})
\frac{1}{\sqrt{aa^{\dagger}\otimes {\bf 1}}}
\sin(\alpha\sqrt{aa^{\dagger}\otimes {\bf 1}})(a\otimes {\bf 1}) \\
&=&-i\ e^{-it\omega_{0}N}\frac{1}{\sqrt{N+1}}\sin(\alpha\sqrt{N+1})a\otimes
e^{it\omega_{0}N}\cos(\alpha\sqrt{N+1}), \\
(14)
&=&e^{-it\omega_{0}K_{0}}
\frac{1}{\sqrt{{\bf 1}\otimes aa^{\dagger}}}
\sin(\alpha\sqrt{{\bf 1}\otimes aa^{\dagger}})({\bf 1}\otimes (a^{\dagger})^{T})
\frac{1}{\sqrt{aa^{\dagger}\otimes {\bf 1}}}
\sin(\alpha\sqrt{aa^{\dagger}\otimes {\bf 1}})(a\otimes {\bf 1}) \\
&=&e^{-it\omega_{0}N}\frac{1}{\sqrt{N+1}}\sin(\alpha\sqrt{N+1})a\otimes
e^{it\omega_{0}N}\frac{1}{\sqrt{N+1}}\sin(\alpha\sqrt{N+1})(a^{\dagger})^{T}; 
\end{eqnarray*}
and

\begin{eqnarray*}
(21)
&=&ie^{-it\omega_{0}}e^{-it\omega_{0}K_{0}}
\frac{1}{\sqrt{{\bf 1}\otimes a^{\dagger}a}}\sin(\alpha\sqrt{{\bf 1}\otimes a^{\dagger}a})
({\bf 1}\otimes a^{T})
\cos(\alpha\sqrt{aa^{\dagger}\otimes {\bf 1}}) \\
&=&ie^{-it\omega_{0}}\ 
e^{-it\omega_{0}N}\cos(\alpha\sqrt{N+1})\otimes
e^{it\omega_{0}N}\frac{1}{\sqrt{N}}\sin(\alpha\sqrt{N})a^{T}, \\
(22)
&=&e^{-it\omega_{0}}e^{-it\omega_{0}K_{0}}
\cos(\alpha\sqrt{{\bf 1}\otimes a^{\dagger}a})
\cos(\alpha\sqrt{aa^{\dagger}\otimes {\bf 1}}) \\
&=&e^{-it\omega_{0}}\ 
e^{-it\omega_{0}N}\cos(\alpha\sqrt{N+1})\otimes
e^{it\omega_{0}N}\cos(\alpha\sqrt{N}), \\
(23)
&=&e^{-it\omega_{0}}e^{-it\omega_{0}K_{0}}
\frac{1}{\sqrt{{\bf 1}\otimes a^{\dagger}a}}\sin(\alpha\sqrt{{\bf 1}\otimes a^{\dagger}a})
({\bf 1}\otimes a^{T})
\frac{1}{\sqrt{aa^{\dagger}\otimes {\bf 1}}}
\sin(\alpha\sqrt{aa^{\dagger}\otimes {\bf 1}})(a\otimes {\bf 1}) \\
&=&e^{-it\omega_{0}}\ 
e^{-it\omega_{0}N}\frac{1}{\sqrt{N+1}}\sin(\alpha\sqrt{N+1})a\otimes
e^{it\omega_{0}N}\frac{1}{\sqrt{N}}\sin(\alpha\sqrt{N})a^{T}, \\
(24)
&=&-ie^{-it\omega_{0}}e^{-it\omega_{0}K_{0}}
\cos(\alpha\sqrt{{\bf 1}\otimes a^{\dagger}a})
\frac{1}{\sqrt{aa^{\dagger}\otimes {\bf 1}}}
\sin(\alpha\sqrt{aa^{\dagger}\otimes {\bf 1}})(a\otimes {\bf 1}) \\
&=&-ie^{-it\omega_{0}}\ 
e^{-it\omega_{0}N}\frac{1}{\sqrt{N+1}}\sin(\alpha\sqrt{N+1})a\otimes
e^{it\omega_{0}N}\cos(\alpha\sqrt{N}); 
\end{eqnarray*}
and
\begin{eqnarray*}
(31)
&=&-ie^{it\omega_{0}}e^{-it\omega_{0}K_{0}}
\cos(\alpha\sqrt{{\bf 1}\otimes aa^{\dagger}})
\frac{1}{\sqrt{a^{\dagger}a\otimes {\bf 1}}}
\sin(\alpha\sqrt{a^{\dagger}a\otimes {\bf 1}})(a^{\dagger}\otimes {\bf 1}) \\
&=&-ie^{it\omega_{0}}\ 
e^{-it\omega_{0}N}\frac{1}{\sqrt{N}}\sin(\alpha\sqrt{N})a^{\dagger}\otimes
e^{it\omega_{0}N}\cos(\alpha\sqrt{N+1}), \\
(32)
&=&e^{it\omega_{0}}e^{-it\omega_{0}K_{0}}
\frac{1}{\sqrt{{\bf 1}\otimes aa^{\dagger}}}
\sin(\alpha\sqrt{{\bf 1}\otimes aa^{\dagger}})({\bf 1}\otimes (a^{\dagger})^{T})
\frac{1}{\sqrt{a^{\dagger}a\otimes {\bf 1}}}
\sin(\alpha\sqrt{a^{\dagger}a\otimes {\bf 1}})(a^{\dagger}\otimes {\bf 1}) \\
&=&e^{it\omega_{0}}\ 
e^{-it\omega_{0}N}\frac{1}{\sqrt{N}}\sin(\alpha\sqrt{N})a^{\dagger}\otimes
e^{it\omega_{0}N}\frac{1}{\sqrt{N+1}}\sin(\alpha\sqrt{N+1})(a^{\dagger})^{T}, \\
(33)
&=&e^{it\omega_{0}}e^{-it\omega_{0}K_{0}}
\cos(\alpha\sqrt{{\bf 1}\otimes aa^{\dagger}})\cos(\alpha \sqrt{a^{\dagger}a\otimes {\bf 1}})
\\
&=&e^{it\omega_{0}}\ 
e^{-it\omega_{0}N}\cos(\alpha\sqrt{N})\otimes 
e^{it\omega_{0}N}\cos(\alpha\sqrt{N+1}), \\
(34)
&=&ie^{it\omega_{0}}e^{-it\omega_{0}K_{0}}
\frac{1}{\sqrt{{\bf 1}\otimes aa^{\dagger}}}
\sin(\alpha\sqrt{{\bf 1}\otimes aa^{\dagger}})({\bf 1}\otimes (a^{\dagger})^{T})
\cos(\alpha \sqrt{a^{\dagger}a\otimes {\bf 1}}) \\
&=&ie^{it\omega_{0}}\ 
e^{-it\omega_{0}N}\cos(\alpha\sqrt{N})\otimes 
e^{it\omega_{0}N}\frac{1}{\sqrt{N+1}}\sin(\alpha\sqrt{N+1})(a^{\dagger})^{T};
\end{eqnarray*}
and

\begin{eqnarray*}
(41)
&=&e^{-it\omega_{0}K_{0}}
\frac{1}{\sqrt{{\bf 1}\otimes a^{\dagger}a}}\sin(\alpha\sqrt{{\bf 1}\otimes a^{\dagger}a})
({\bf 1}\otimes a^{T})
\frac{1}{\sqrt{a^{\dagger}a\otimes {\bf 1}}}
\sin(\alpha\sqrt{a^{\dagger}a\otimes {\bf 1}})(a^{\dagger}\otimes {\bf 1}) \\
&=&
e^{-it\omega_{0}N}\frac{1}{\sqrt{N}}\sin(\alpha\sqrt{N})a^{\dagger}\otimes
e^{it\omega_{0}N}\frac{1}{\sqrt{N}}\sin(\alpha\sqrt{N})a^{T}, \\
(42)
&=&-ie^{-it\omega_{0}K_{0}}
\cos(\alpha\sqrt{{\bf 1}\otimes a^{\dagger}a})
\frac{1}{\sqrt{a^{\dagger}a\otimes {\bf 1}}}
\sin(\alpha\sqrt{a^{\dagger}a\otimes {\bf 1}})(a^{\dagger}\otimes {\bf 1}) \\
&=&-i\ 
e^{-it\omega_{0}N}\frac{1}{\sqrt{N}}\sin(\alpha\sqrt{N})a^{\dagger}\otimes
e^{it\omega_{0}N}\cos(\alpha\sqrt{N}), \\
(43)
&=&ie^{-it\omega_{0}K_{0}}
\frac{1}{\sqrt{{\bf 1}\otimes a^{\dagger}a}}\sin(\alpha\sqrt{{\bf 1}\otimes a^{\dagger}a})
({\bf 1}\otimes a^{T})\cos(\alpha \sqrt{a^{\dagger}a\otimes {\bf 1}}) \\
&=&i\ 
e^{-it\omega_{0}N}\cos(\alpha\sqrt{N})\otimes
e^{it\omega_{0}N}\frac{1}{\sqrt{N}}\sin(\alpha\sqrt{N})a^{T}, \\
(44)
&=&e^{-it\omega_{0}K_{0}}
\cos(\alpha\sqrt{{\bf 1}\otimes a^{\dagger}a})\cos(\alpha \sqrt{a^{\dagger}a\otimes {\bf 1}})
\\
&=&
e^{-it\omega_{0}N}\cos(\alpha\sqrt{N})\otimes
e^{it\omega_{0}N}\cos(\alpha\sqrt{N}).
\end{eqnarray*}

\vspace{5mm}\noindent
[II]\ \ Second, we must calculate $e^{tY}$. To be glad this task has been 
done, see \cite{EFS} and \cite{FS}. 
Namely, from (\ref{eq:Y modified}) we have
\begin{equation}
\label{eq:II}
e^{tY}=
\left(
\begin{array}{cccc}
e^{tL} &          &          &          \\
         & e^{tL} &          &          \\
         &          & e^{tL} &          \\
         &          &          & e^{tL}
\end{array}
\right);
\quad 
L=\frac{\mu-\nu}{2}+\nu K_{+}+\mu K_{-}-(\mu+\nu)K_{3}
\end{equation}
and
\begin{eqnarray}
e^{tL}
&=&
e^{\frac{\mu-\nu}{2}t}e^{t\{\nu K_{+}+\mu K_{-}-(\mu+\nu)K_{3}\}} \nonumber \\
&=&
e^{\frac{\mu-\nu}{2}t}
e^{G(t)K_{+}}e^{-2\log(F(t))K_{3}}e^{E(t)K_{-}}
\end{eqnarray}
with
\begin{eqnarray}
\label{eq:fundamental functions}
E(t)&=&\frac{\frac{2\mu}{\mu-\nu}\sinh\left(\frac{\mu-\nu}{2}t\right)}
     {\cosh\left(\frac{\mu-\nu}{2}t\right)+\frac{\mu+\nu}{\mu-\nu}
      \sinh\left(\frac{\mu-\nu}{2}t\right)}, \nonumber \\
F(t)&=&\cosh\left(\frac{\mu-\nu}{2}t\right)+
     \frac{\mu+\nu}{\mu-\nu}\sinh\left(\frac{\mu-\nu}{2}t\right), 
\nonumber \\
G(t)&=&\frac{\frac{2\nu}{\mu-\nu}\sinh\left(\frac{\mu-\nu}{2}t\right)}
     {\cosh\left(\frac{\mu-\nu}{2}t\right)+\frac{\mu+\nu}{\mu-\nu}
      \sinh\left(\frac{\mu-\nu}{2}t\right)}.
\end{eqnarray}
This is a kind of disentangling formula, see for example \cite{KF} 
as a general introduction.

\vspace{5mm}\noindent
[III]\ \ Third, we must calculate $e^{-i\frac{t^{2}}{2}[X,Y]}$. 
In this stage some interaction appears. Let us calculate 
$[X,Y]$ exactly.

Some calculation gives
\begin{eqnarray}
&&[{\bf 1}\otimes a^{T}, \nu K_{+}+\mu K_{-}-(\mu+\nu)K_{3}]
=
-\mu a\otimes {\bf 1}+\frac{\mu+\nu}{2} {\bf 1}\otimes a^{T}, 
\nonumber \\
&&[{\bf 1}\otimes (a^{\dagger})^{T}, \nu K_{+}+\mu K_{-}-(\mu+\nu)K_{3}]
=
\nu a^{\dagger}\otimes {\bf 1}-\frac{\mu+\nu}{2} {\bf 1}\otimes (a^{\dagger})^{T}, 
\nonumber \\
&&[a\otimes {\bf 1}, \nu K_{+}+\mu K_{-}-(\mu+\nu)K_{3}]
=
\nu {\bf 1}\otimes a^{T}-\frac{\mu+\nu}{2} a\otimes {\bf 1}, 
\nonumber \\
&&[a^{\dagger}\otimes {\bf 1}, \nu K_{+}+\mu K_{-}-(\mu+\nu)K_{3}]
=
-\mu {\bf 1}\otimes (a^{\dagger})^{T}+\frac{\mu+\nu}{2} a^{\dagger}\otimes {\bf 1}, 
\nonumber \\
&&[K_{0},\nu K_{+}+\mu K_{-}-(\mu+\nu)K_{3}]=0
\end{eqnarray}
and for simplicity we set
\begin{eqnarray}
A&=&-\nu a^{\dagger}\otimes {\bf 1}+\frac{\mu+\nu}{2} {\bf 1}\otimes (a^{\dagger})^{T}, 
\nonumber \\
B&=&\mu a\otimes {\bf 1}-\frac{\mu+\nu}{2} {\bf 1}\otimes a^{T}, 
\nonumber \\
C&=&\nu {\bf 1}\otimes a^{T}-\frac{\mu+\nu}{2} a\otimes {\bf 1}, 
\nonumber \\
D&=&-\mu {\bf 1}\otimes (a^{\dagger})^{T}+\frac{\mu+\nu}{2} a^{\dagger}\otimes {\bf 1}.
\end{eqnarray}
Note that $B\ne -A^{\dagger}$ and $D\ne -C^{\dagger}$ because of 
$\mu > \nu$. It is easy to see
\begin{eqnarray}
\label{eq:relations III}
&&[A,C]=[A,D]=0,\quad [B,C]=[B,D]=0, \nonumber \\
&&[A,B]=[C,D]=-\left(\frac{\mu-\nu}{2}\right)^{2}.
\end{eqnarray}

From the decomposition (\ref{eq:decomposition I}) we have a decomposition 
consisting of two commutative operators
\begin{equation}
[X,Y]
=
\Omega
\left(
\begin{array}{cccc}
0 & A & 0 & 0   \\
B & 0 & 0 & 0   \\
0 & 0 & 0 & A   \\
0 & 0 & B & 0
\end{array}
\right)
+
\Omega
S
\left(
\begin{array}{cccc}
0 & C & 0 & 0   \\
D & 0 & 0 & 0   \\
0 & 0 & 0 & C   \\
0 & 0 & D & 0
\end{array}
\right)
S
\end{equation}
with $S$ in (\ref{eq:swap}).

Now, by making use of the formula
\begin{equation}
\mbox{exp}
\left(
-i\beta
\left(
\begin{array}{cc}
  0 & P  \\
  Q & 0
\end{array}
\right)
\right)
=
\left(
\begin{array}{cc}
\cos(\beta\sqrt{PQ}) & -i\frac{1}{\sqrt{PQ}}\sin(\beta\sqrt{PQ})P  \\
-i\frac{1}{\sqrt{QP}}\sin(\beta\sqrt{QP})Q & \cos(\beta\sqrt{QP}) 
\end{array}
\right)
\end{equation}
for some parameter $\beta$, we can calculate the term 
$e^{-i\frac{t^{2}}{2}[X,Y]}$ easily. The result is 
\begin{equation}
\label{eq:III}
e^{-i\frac{t^{2}}{2}[X,Y]}
=
\left(
\begin{array}{cccc}
  (11) & (12) & (13) & (14)  \\
  (21) & (22) & (23) & (24)  \\
  (31) & (32) & (33) & (34)  \\
  (41) & (42) & (43) & (44)
\end{array}
\right)
\end{equation}
where for $\beta=\Omega\frac{t^{2}}{2}$
\begin{eqnarray}
&&(11)=\cos(\beta\sqrt{AB})\cos(\beta\sqrt{CD}), 
\nonumber \\
&&(12)=-i\frac{1}{\sqrt{AB}}\sin(\beta\sqrt{AB})A\cos(\beta\sqrt{CD}), 
\nonumber \\
&&(13)=-i\cos(\beta\sqrt{AB})\frac{1}{\sqrt{CD}}\sin(\beta\sqrt{CD})C, 
\nonumber \\
&&(14)=-\frac{1}{\sqrt{AB}}\sin(\beta\sqrt{AB})A\frac{1}{\sqrt{CD}}\sin(\beta\sqrt{CD})C\ ;
\nonumber \\
&{}& \nonumber \\
&&(21)=-i\frac{1}{\sqrt{BA}}\sin(\beta\sqrt{BA})B\cos(\beta\sqrt{CD}), 
\nonumber \\
&&(22)=\cos(\beta\sqrt{BA})\cos(\beta\sqrt{CD}), 
\nonumber \\
&&(23)=-\frac{1}{\sqrt{BA}}\sin(\beta\sqrt{BA})B\frac{1}{\sqrt{CD}}\sin(\beta\sqrt{CD})C, 
\nonumber \\
&&(24)=-i\cos(\beta\sqrt{BA})\frac{1}{\sqrt{CD}}\sin(\beta\sqrt{CD})C\ ;
\nonumber \\
&{}& \nonumber \\
&&(31)=-i\cos(\beta\sqrt{AB})\frac{1}{\sqrt{DC}}\sin(\beta\sqrt{DC})D,
\nonumber \\
&&(32)=-\frac{1}{\sqrt{AB}}\sin(\beta\sqrt{AB})A\frac{1}{\sqrt{DC}}\sin(\beta\sqrt{DC})D,
\nonumber \\
&&(33)=\cos(\beta\sqrt{AB})\cos(\beta\sqrt{DC}), 
\nonumber \\
&&(34)=-i\frac{1}{\sqrt{AB}}\sin(\beta\sqrt{AB})A\cos(\beta\sqrt{DC})\ ;
\nonumber
\end{eqnarray}
\begin{eqnarray}
&{}& \nonumber \\
&&(41)=-\frac{1}{\sqrt{BA}}\sin(\beta\sqrt{BA})B\frac{1}{\sqrt{DC}}\sin(\beta\sqrt{DC})D,
\nonumber \\
&&(42)=-i\cos(\beta\sqrt{BA})\frac{1}{\sqrt{DC}}\sin(\beta\sqrt{DC})D,
\nonumber \\
&&(43)=-i\frac{1}{\sqrt{BA}}\sin(\beta\sqrt{BA})B\cos(\beta\sqrt{DC}),
\nonumber \\
&&(44)=\cos(\beta\sqrt{BA})\cos(\beta\sqrt{DC}) \nonumber
\end{eqnarray}
where
\begin{eqnarray*}
AB=
-\mu\nu a^{\dagger}a\otimes {\bf 1}
+\frac{\mu+\nu}{2}\nu a^{\dagger}\otimes a^{T}
+\mu\frac{\mu+\nu}{2} a\otimes (a^{\dagger})^{T}
-\left(\frac{\mu+\nu}{2}\right)^{2} {\bf 1}\otimes aa^{\dagger}, \\
BA=
-\mu\nu aa^{\dagger}\otimes {\bf 1}
+\mu\frac{\mu+\nu}{2} a\otimes (a^{\dagger})^{T}
+\frac{\mu+\nu}{2}\nu a^{\dagger}\otimes a^{T}
-\left(\frac{\mu+\nu}{2}\right)^{2} {\bf 1}\otimes a^{\dagger}a, \\
CD=
-\mu\nu {\bf 1}\otimes a^{\dagger}a
+\frac{\mu+\nu}{2}\nu a^{\dagger}\otimes a^{T}
+\mu\frac{\mu+\nu}{2} a\otimes (a^{\dagger})^{T}
-\left(\frac{\mu+\nu}{2}\right)^{2} aa^{\dagger}\otimes {\bf 1}, \\
DC=
-\mu\nu {\bf 1}\otimes aa^{\dagger}
+\mu\frac{\mu+\nu}{2} a\otimes (a^{\dagger})^{T}
+\frac{\mu+\nu}{2}\nu a^{\dagger}\otimes a^{T}
-\left(\frac{\mu+\nu}{2}\right)^{2} a^{\dagger}a\otimes {\bf 1}.
\end{eqnarray*}
These are complicated enough.

Anyway, we have completed the task although it is very complicated.

\vspace{5mm}
In last, we shall restore the result to original form. If we set
\begin{equation}
\label{eq:approximate formula}
\hat{\tilde{\rho}}(t)=e^{tY}e^{-itX}\hat{\rho}(0)
\equiv e^{tY}\hat{\tilde{\rho}}_{1}(t),
\end{equation}
then $\tilde{\rho}_{1}(t)$ is from (\ref{eq:I}) given by
\begin{equation}
\tilde{\rho}_{1}(t)
=
\left(
\begin{array}{cc}
(11) & (12) \\
(21) & (22)
\end{array}
\right)
\end{equation}
where

\begin{eqnarray*}
(11)
&=&
e^{-it\omega_{0}N}\cos(\alpha\sqrt{N+1})\ {\rho_{00}}\
e^{it\omega_{0}N}\cos(\alpha\sqrt{N+1}) \\
&{}&+ie^{-it\omega_{0}N}\cos(\alpha\sqrt{N+1})\ {\rho_{01}}\  
a^{\dagger}e^{it\omega_{0}N}\frac{1}{\sqrt{N+1}}\sin(\alpha\sqrt{N+1}) \\
&{}&-ie^{-it\omega_{0}N}\frac{1}{\sqrt{N+1}}\sin(\alpha\sqrt{N+1})a\ {\rho_{10}}\ 
e^{it\omega_{0}N}\cos(\alpha\sqrt{N+1}) \\
&{}&+e^{-it\omega_{0}N}\frac{1}{\sqrt{N+1}}\sin(\alpha\sqrt{N+1})a\ {\rho_{11}}\ 
a^{\dagger}e^{it\omega_{0}N}\frac{1}{\sqrt{N+1}}\sin(\alpha\sqrt{N+1}), \\
(12)
&=&ie^{-it\omega_{0}}\ 
e^{-it\omega_{0}N}\cos(\alpha\sqrt{N+1})\ {\rho_{00}}\
a e^{it\omega_{0}N}\frac{1}{\sqrt{N}}\sin(\alpha\sqrt{N}) \\
&{}&+e^{-it\omega_{0}}\
e^{-it\omega_{0}N}\cos(\alpha\sqrt{N+1})\ {\rho_{01}}\  
e^{it\omega_{0}N}\cos(\alpha\sqrt{N}) \\
&{}&+e^{-it\omega_{0}}\
e^{-it\omega_{0}N}\frac{1}{\sqrt{N+1}}\sin(\alpha\sqrt{N+1})a\ {\rho_{10}}\ 
a e^{it\omega_{0}N}\frac{1}{\sqrt{N}}\sin(\alpha\sqrt{N}) \\
&{}&-ie^{-it\omega_{0}}\
e^{-it\omega_{0}N}\frac{1}{\sqrt{N+1}}\sin(\alpha\sqrt{N+1})a\ {\rho_{11}}\ 
e^{it\omega_{0}N}\cos(\alpha\sqrt{N}), \\
(21)
&=&-ie^{it\omega_{0}}\
e^{-it\omega_{0}N}\frac{1}{\sqrt{N}}\sin(\alpha\sqrt{N})a^{\dagger}\ {\rho_{00}}\
e^{it\omega_{0}N}\cos(\alpha\sqrt{N+1}) \\
&{}&+e^{it\omega_{0}}\
e^{-it\omega_{0}N}\frac{1}{\sqrt{N}}\sin(\alpha\sqrt{N})a^{\dagger}\ {\rho_{01}}\
a^{\dagger}e^{it\omega_{0}N}\frac{1}{\sqrt{N+1}}\sin(\alpha\sqrt{N+1}) \\
&{}&+e^{it\omega_{0}}\
e^{-it\omega_{0}N}\cos(\alpha\sqrt{N})\ {\rho_{10}}\ 
e^{it\omega_{0}N}\cos(\alpha\sqrt{N+1}) \\
&{}&+ie^{it\omega_{0}}\
e^{-it\omega_{0}N}\cos(\alpha\sqrt{N})\ {\rho_{11}}\
a^{\dagger}e^{it\omega_{0}N}\frac{1}{\sqrt{N+1}}\sin(\alpha\sqrt{N+1}), \\
(22)
&=&e^{-it\omega_{0}N}\frac{1}{\sqrt{N}}\sin(\alpha\sqrt{N})a^{\dagger}\ {\rho_{00}}\
ae^{it\omega_{0}N}\frac{1}{\sqrt{N}}\sin(\alpha\sqrt{N}) \\
&{}&
-ie^{-it\omega_{0}N}\frac{1}{\sqrt{N}}\sin(\alpha\sqrt{N})a^{\dagger}\ {\rho_{01}}\
e^{it\omega_{0}N}\cos(\alpha\sqrt{N}) \\
&{}&
+ie^{-it\omega_{0}N}\cos(\alpha\sqrt{N})\ {\rho_{10}}\
ae^{it\omega_{0}N}\frac{1}{\sqrt{N}}\sin(\alpha\sqrt{N}) \\
&{}&
+e^{-it\omega_{0}N}\cos(\alpha\sqrt{N})\ {\rho_{11}}\
e^{it\omega_{0}N}\cos(\alpha\sqrt{N}),
\end{eqnarray*}
and $\tilde{\rho}(t)$ is by
\begin{eqnarray}
\label{eq:standard form}
\tilde{\rho}(t)
&=&
\frac{\mbox{e}^{\frac{\mu-\nu}{2}t}}{F(t)}
\sum_{n=0}^{\infty}
\frac{G(t)^{n}}{n!}(a^{\dagger})^{n}
[
\exp\left(\{-\log(F(t))\}N\right)
\left\{
\sum_{m=0}^{\infty}
\frac{E(t)^{m}}{m!}a^{m}\tilde{\rho}_{1}(t)(a^{\dagger})^{m}
\right\}\times \nonumber \\
&&
\qquad\qquad \exp\left(\{-\log(F(t))\}N\right)
]
a^{n}
\end{eqnarray}
in terms of $E(t), F(t), G(t)$ in (\ref{eq:fundamental functions}) 
and $N=a^{\dagger}a$. See \cite{EFS} and \cite{FS}. 
This form is relatively clear because of no interaction.

Our (approximate) solution is
\[
\hat{\rho}(t)
=e^{-i\frac{t^{2}}{2}[X,Y]}e^{tY}e^{-itX}
=e^{-i\frac{t^{2}}{2}[X,Y]}\hat{\tilde{\rho}}(t)
\]
from (\ref{eq:approximate formula}). 
We would like to express $\rho(t)$ like (\ref{eq:standard form}). 
From (\ref{eq:III}) we can expand each term in terms of the Taylor 
expansion of $\cos(X)$ and $\sin(X)$.  
However, the form is very ugly in the Dirac's sense, so we leave 
such a task to readers.

\vspace{10mm}\noindent
{\bf Concluding Remarks}\ \ In this paper we treated the master equation 
for the density operator based on the Jaynes--Cummings Hamiltonian 
with dissipation and constructed the approximate solution up to $O(t^{3})$ 
under the general setting. Our construction is quite general because 
$\hat{\rho}(0)$ is any initial state.

However, it may be still inconvenient to apply to realistic problems 
coming from the atom--cavity system. Further work will be required.

Moreover, it may be possible to treat a generalized master equation for 
the density operator given by
\begin{eqnarray}
\label{eq:quantum damped Jaynes-Cummings II}
\frac{\partial}{\partial t}\rho
&=&-i[H_{JC},\rho]
+
{\mu}
\left\{a\rho a^{\dagger}-\frac{1}{2}(a^{\dagger}a\rho+\rho a^{\dagger}a)\right\}
+
{\nu}
\left\{a^{\dagger}\rho{a}-\frac{1}{2}(aa^{\dagger}\rho+\rho aa^{\dagger})\right\}
+
\nonumber \\
&&
{\kappa}\left\{a\rho a-\frac{1}{2}(a^{2}\rho+\rho a^{2})\right\}
+
{\bar{\kappa}}\left\{a^{\dagger}\rho a^{\dagger}
-\frac{1}{2}((a^{\dagger})^{2}\rho+\rho (a^{\dagger})^{2})\right\}
\end{eqnarray}
with the positivity condition ${\mu}{\nu}\geq |\kappa|^{2}$.  
However, we don't treat this general case in the paper, see \cite{KF2}, 
\cite{KF3} and \cite{KF4}.

\vspace{10mm}

\end{document}